\documentclass[a4paper,
               keeplastbox,   
               ]{jacow}
\usepackage{pdfpages,multirow,ragged2e} %
\makeatletter%
	\ifboolexpr{bool{xetex}}
	 {\renewcommand{\Gin@extensions}{.pdf,%
	                    .png,.jpg,.bmp,.pict,.tif,.psd,.mac,.sga,.tga,.gif,%
	                    .eps,.ps,%
	                    }}{}
\makeatother

\ifboolexpr{bool{xetex} or bool{luatex}} 
 {}                                      
 {\usepackage[utf8]{inputenc}}           

\usepackage[USenglish]{babel}
\usepackage{subfigure}
\usepackage{epstopdf}
\usepackage{bm}
\usepackage{dcolumn}
\usepackage{subfloat}
\usepackage{booktabs}
\usepackage{float}
\usepackage{epstopdf}
\usepackage{gensymb}

\ifboolexpr{bool{jacowbiblatex}}%
 {%
  \addbibresource{jacow-test.bib}
  \addbibresource{biblatex-examples.bib}
 }{}
\begin{document}

\title{MANY-OBJECTIVE BEAM DYNAMICS OPTIMIZATION FOR HIGH-REPETITION-RATE XFEL PHOTOINJECTOR}

\author{Z. H. Zhu\textsuperscript{1,2}, J. W. Yan\textsuperscript{1,2}, D. Gu\textsuperscript{3}, Q. Gu\textsuperscript{3}\thanks{guqiang@zjlab.org.cn}, \\
\textsuperscript{1}Shanghai Institute of Applied Physics, Chinese Academy of Sciences, Shanghai 201800, China\\
		\textsuperscript{2}University of Chinese Academy of Sciences, Beijing 100049, China\\
	\textsuperscript{3}Shanghai Advanced Research Institute, Chinese Academy of Sciences, Shanghai 201210, China}
	
\maketitle

\begin{abstract}
SHINE, as the first hard x-ray free-electron-laser (FEL) facility in China, is designed to provide high-brightness FEL lasing under high-repetition-rate operation mode. In order to drive x-ray FEL pulses with high qualities, the photoinjector section is deployed to provide the specified electron beam with low transverse emittance and high brightness. Normally the multi-objective optimization algorithm is employed in the injector beam dynamics design. In this paper, the many-objective optimization algorithm NSGA-III is introduced to the injector physical design for optimizing the 4 beam quality properties for the first time. The results of the optimization are presented and analyzed. The application of NSGA-III can provide guidance for further physical research as well as improve the beam dynamics optimization efficiency.

\end{abstract}

\section{INTRODUCTION}

Shanghai High repetitioN rate XFEL and Extreme light  facility (SHINE)\cite{SHINE}, which is designed to deliver photon energy from 0.4 – 25 keV at repetition rates as high as 1 MHz, is under construction. The electron bunch is accelerated to 8 GeV through the continuous-wave superconducting linear accelerator. To generate free-electron-laser (FEL) pulses with high brightness, high-energy electron beams with low emittance and short duration are required. 
\begin{table}[!htb]
	\renewcommand\arraystretch{1.5}
	\centering
	\setlength{\tabcolsep}{3.5mm}
	\caption{\label{tabfourpara}%
		}
	\begin{tabular}{cccccccc}
		\hline 
		\textrm{$\rm Parameters      $}&
		\textrm{$\rm Values$}\\
		\hline 
		Energy (GeV)     & 8    \\ 
		Slice energy spread (rms)  &   $0.01 \% $ \\
		Slice emittance (mm·mrad,rms)  & 0.3   \\
		Peak current (A)  & 1500  \\ 
		\hline
	\end{tabular}
\end{table} 
	
Tab.~\ \ref{tabfourpara} shows the required beam properites at the end of linac for SHINE FEL performance. The transverse emittance with a value of lower than 0.3 is required, which is mainly determined in the photoinjector section. In addition, the peak current of 1.5 kA is achieved after the two bunch compression chicanes in the main linac section. Nevertheless, strong nonlinear effects are often generated in bunch longitudinal compression, leading to the sharp horns in the current profile. These current horns are undesired for inducing the coherent synchrotron radiation (CSR) effect, spoiling the bunch transverse emittance and leading to the unwanted energy modulation in the longitudinal phase space \cite{csr}. 
	
\begin{figure}[htp] 
	\centering 
	\includegraphics[width=1\linewidth]{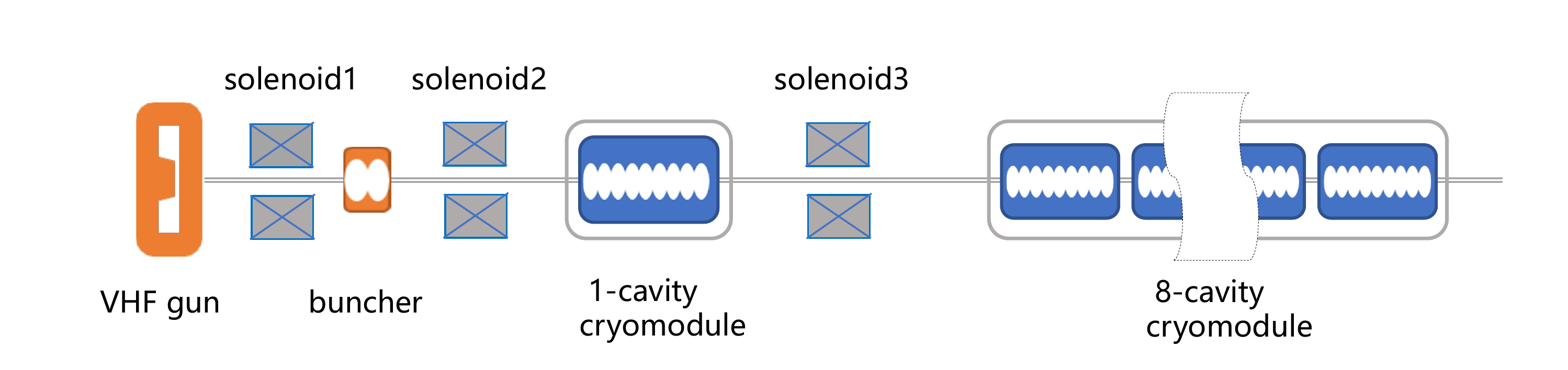} 
	\caption{Schematic layout of SHINE injector section.The beam is accelerated to 100 MeV with peak current of 10A at the exit of the injector.}
	\label{layout} 
\end{figure}

The schematic of the photoinjector section is shown in FIG.\ \ref{layout}. This system consists of the electron gun operating at a CW mode at the very-high-frequency band. The electron is launched from the photocathode and accelerated to about 0.8 MeV at the exit of the gun. The longer driven laser pulse is settled at the cathode for relieving the space charge effect which will dominate until the single-cavity cryomodule due to the relatively low gradient in the VHF gun. The buncher is followed by the gun exit and compresses the longitudinal distribution of the bunch utilizing the velocity difference along the bunch. The distribution modulation increases the peak current to about 10A at the entrance of the main accelerating section. In addition, three sets of solenoids are inserted to compensate for the linear emittance growth resulted from the space charge force and RF field. However, the bunch often suffers from nonlinear compression in the chicanes, which leads to current spike formation and/or decreases the current of the core part of the bunch.

\section{Beam dynamics optimization}
The main sources of the nonlinear compression come from the nonlinear energy chirp and high-order terms of the dispersion coefficient in the chicane. The nonlinearity in the correlated energy chirp mainly results from the RF sinusoidal curvature, wakefields effect, and space charge effect during the low $\beta$ beam transporting range \cite{mitri}. Since the third harmonic cavity is deployed in the main accelerating section for removing the quadratic term of the correlated energy spread \cite{emma}, remaining the higher terms uncompensated, which degrades the strong longitudinal compression, hence the third and higher terms should be evaluated accurately. Also, the cubic and higher terms of correlated energy spread originates from the strong space charge force and nonlinear longitudinal density modulation in the VHF gun and the buncher cavity. Thus, the beam dynamics design and optimization in the injector section is significant for the bunch magnetic compression performance, both the transverse and longitudinal beam quality should be studied precisely in this space-charge dominated section.

Usually, the transverse emittance and bunch length are treated as the two optimization objective in the injector physical optimization \cite{nsga21,nsga22}. However, in the CW high-repetition-rate XFEL  injector, more detailed beam properties should be considered for precise evaluation because the strong compression in the magnetic chicane is susceptible to the accumulated nonlinearity from the injector. In order to evaluate the nonlinerity of the beam longitudinal phase space, more beam dynamics properties should be treated as the optimization objectives since the bunch length cannot reflect the detailed beam longitudinal phase space distribution. 

The high-order energy spread, which is identified as the main cause of the current horn formation, cannot be mitigated and needs to be minimized. The other beam quality is current profile shape. In this optimization, the skewness is introduced to evaluate the degree of deviation from the symmetric longitudinal distribution of the current profile. In the discipline of probability and statistics, the definition of skewness is:
\begin{equation}
	K=E[(\frac{X-\mu}{\sigma})^{3}]=\frac{\mu_3}{\sigma_3}=\frac{E[(X-\mu)^3]}{(E[(X-\mu)^2)])^{3/2}},
\end{equation}
where X is the random distribution, $\mu_3$ is the third order central moment of the electron distribution, $\sigma$ is the standard deviation and E is the expectation operator. The ideal beam longitudinal distribution is gaussian-like, however, under the influence of strong space charge force and ballistic compression, the beam experiences the nonlinear density modulation and ends up with the asymmetric profile which brings undeired energy modulation by wakefields effect. Together with the emittance and bunch length, four beam properties are selected as the objectives in this optimization, which becomes a many-objective optimization problem. Therefore, the Non-dominated Sorting Genetic Algorithm III (NSGA-III) \cite{NSGA3}, which is an evolutionary many-objective optimization algorithm of which the non-dominated sorting approach is based on reference-point, is selected in this injector optimization. The generation is set to be 150 with population number of 200 in each generation. The beam dynamics simulation is conducted using the ASTRA. 
	
\begin{figure}[htp] 
	\centering 
	\includegraphics[width=1\linewidth]{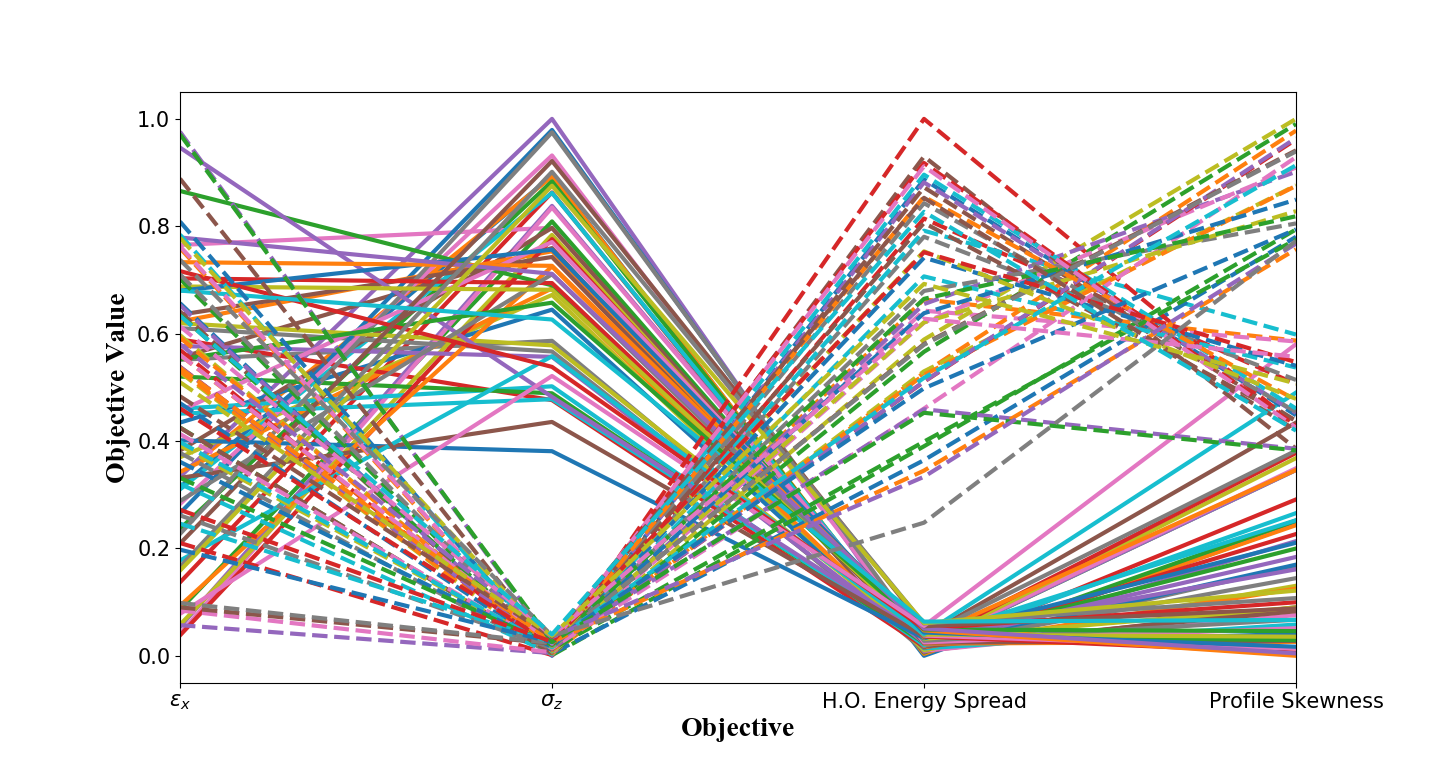}
	\caption{The parallel coordinate plots of the 100 solutions in the final population of generation showing the objectives value path. 50 of them with the shortest bunch length (dashed line) and the other 50 lines present the solutions with the smallest high-order energy spread (solid line).}
	\label{valuepath} 
\end{figure}

\begin{figure}[htb] 
	\centering 
	\subfigure[]{
		\label{project1}	
		\includegraphics[width=0.45\linewidth]{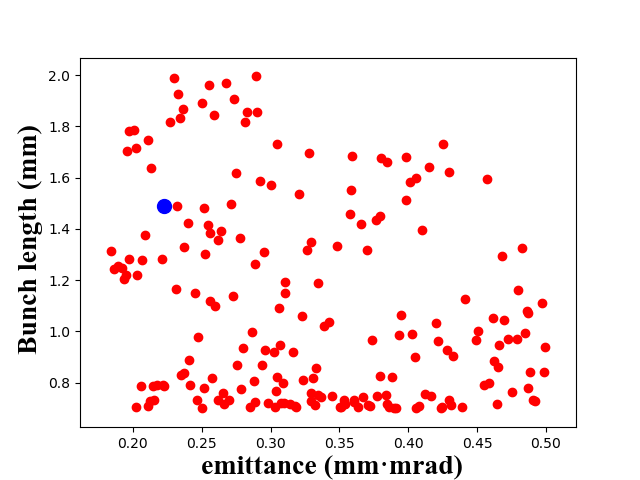}}
	\subfigure[]{
		\label{project2}	
		\includegraphics[width=0.45\linewidth]{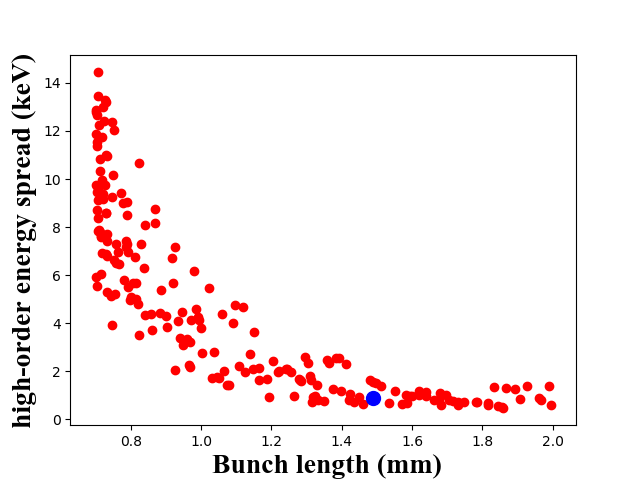}}
	\subfigure[]{
		\label{project3}	
		\includegraphics[width=0.45\linewidth]{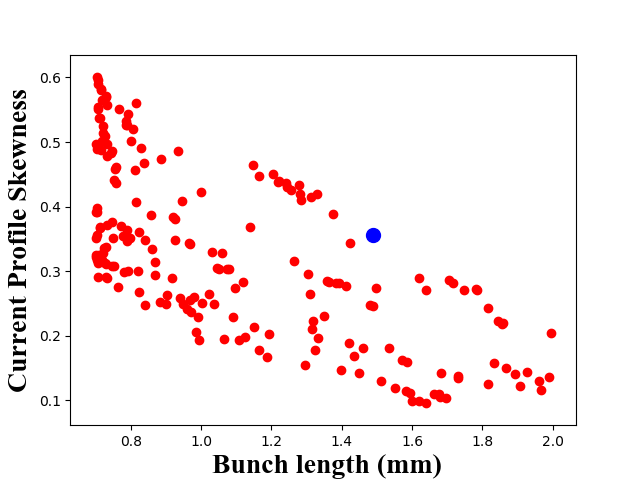}}
	\subfigure[]{
		\label{project4}	
		\includegraphics[width=0.45\linewidth]{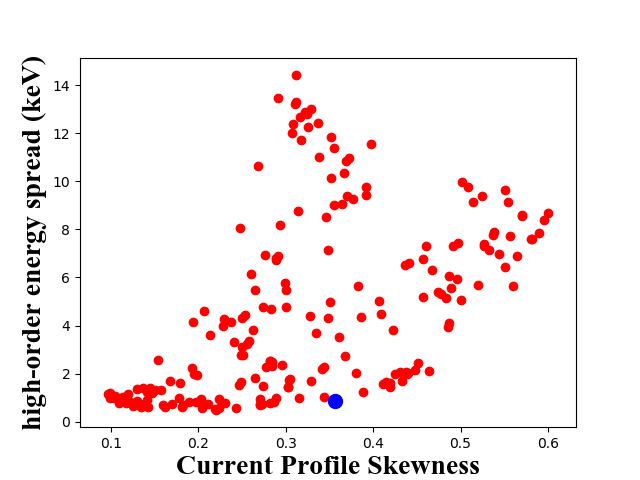}}
	\caption[width=1\textwidth]{The projection of the Pareto front obtained from the last generation. The blue dots in each figure represent the selected solution of which the detailed beam dynamics paramters are shown in FIG.\ \ref{property}}
	\label{project} 
\end{figure}
The optimization results are shown in FIG.\ \ref{valuepath}, which presents the 50 solutions with the shortest bunch length and the other 50 solutions with the smallest high-order energy spread, both the two beam properties are usually selected as the optimization objectives in injector design. In addition, normalization is conducted to all the objective values of the solutions, it is defined as follows:
\begin{equation}
	X_{nij}=\frac{X_{ij}-min(X_{i})}{max(X_{i})-min(X_{i})}
\end{equation}
where $X_{nij}$ is the normalized value, $X_{ij}$ is the $i$th objective of the $j$th solution, $max(X_{i})$ is the maximum value of the $i$th objective among the whole population, $min(X_{i})$ is the minimum value of the $i$th objective among the solutions. Based on this, some distinct corelation can be found in the objectives.

In order to further analyze the results and choose the desired solution to the physical design of the injector, the projections of the Pareto-optimal front are plotted in FIG.\ \ref{project}. FIG.\ \ref{project1} shows the correlation between the transverse normalized emittance and bunch length, unlike the usual tradeoff in the two-objective optimization, there is no clear negative correlation, which indicates that there is much potential relationship between the detailed beam properties in the longitudinal phase space. In FIG.\ \ref{project2}, the bunch length has a great impact on the high-order energy spread. This strong correlation results mainly from the nonlinear longitudinal density modulation in the buncher, the stronger velocity compression in the buncher, the shorter bunch length is achieved at the end of the injector, together with more quadratic and cubic energy modulation imprinted on the beam. In addition, the current profile, which indicates the longitudinal density distribution along the bunch, is also implicated in the nonlinear effect. FIG.\ \ref{project3} and FIG.\ \ref{project4} shows the current profile tends to become asymmetric that not a gaussian-like as desired. If the bunch length is relatively long enough, the high-order energy spread can be ultra-small and the profile symmetry can keep almost gaussian, however, the bunch length is too long to get the desired peak current after the magnetic compression. 

\begin{figure}[htb] 
	\centering 
	\subfigure[]{
		\label{project1}	
		\includegraphics[width=0.48\linewidth]{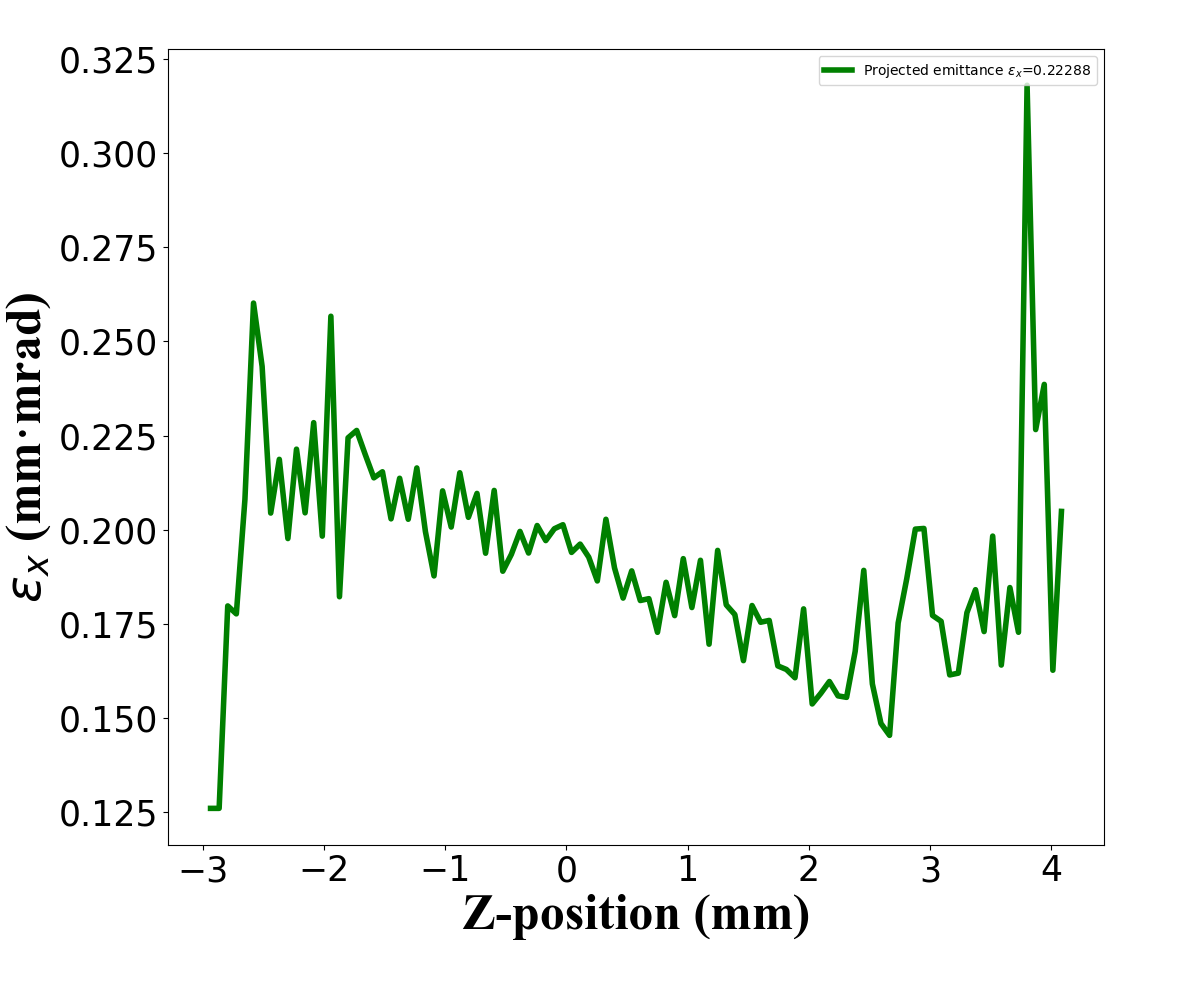}}
	\subfigure[]{
		\label{project2}	
		\includegraphics[width=0.48\linewidth]{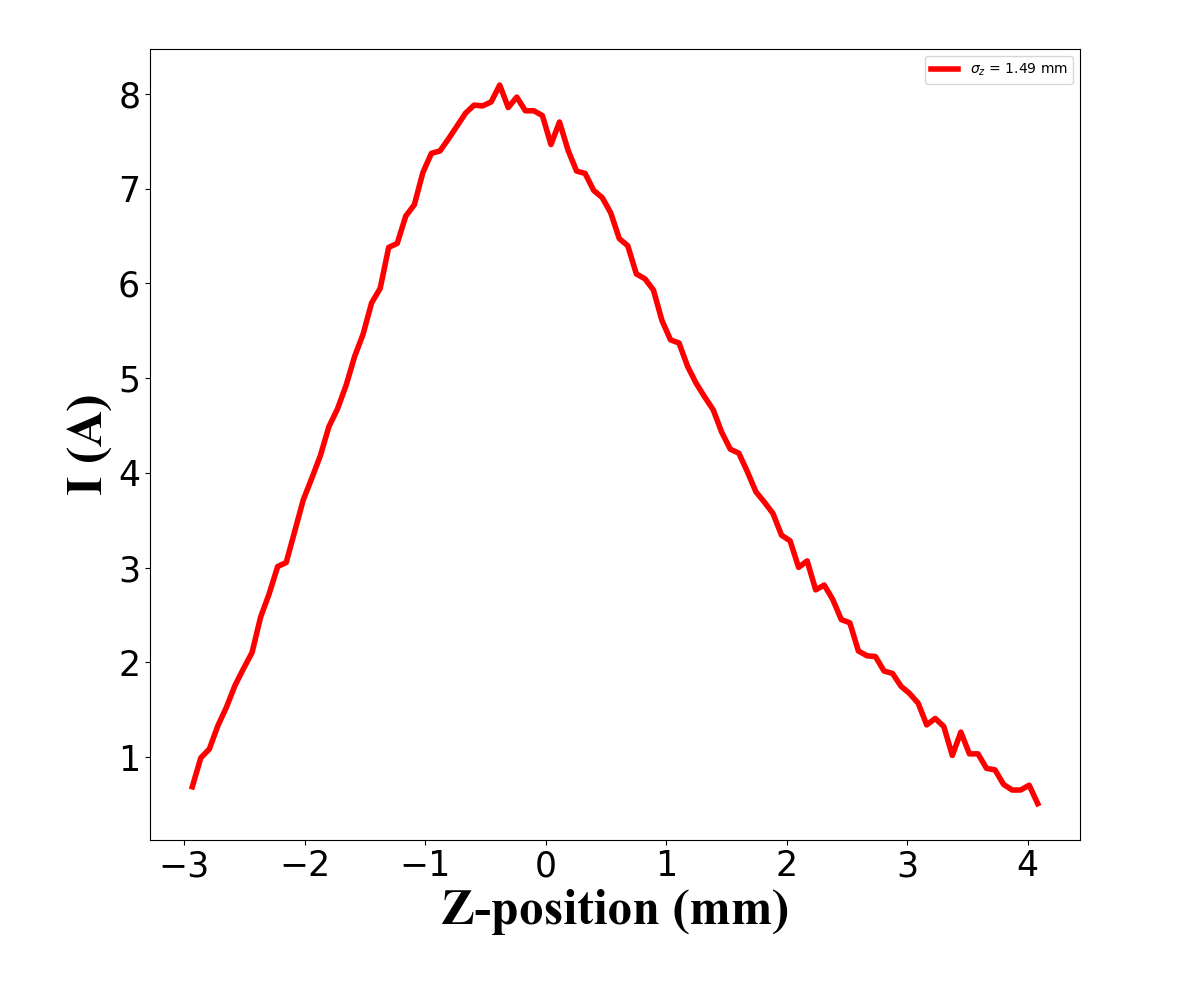}}
	\subfigure[]{
		\label{project3}	
		\includegraphics[width=0.48\linewidth]{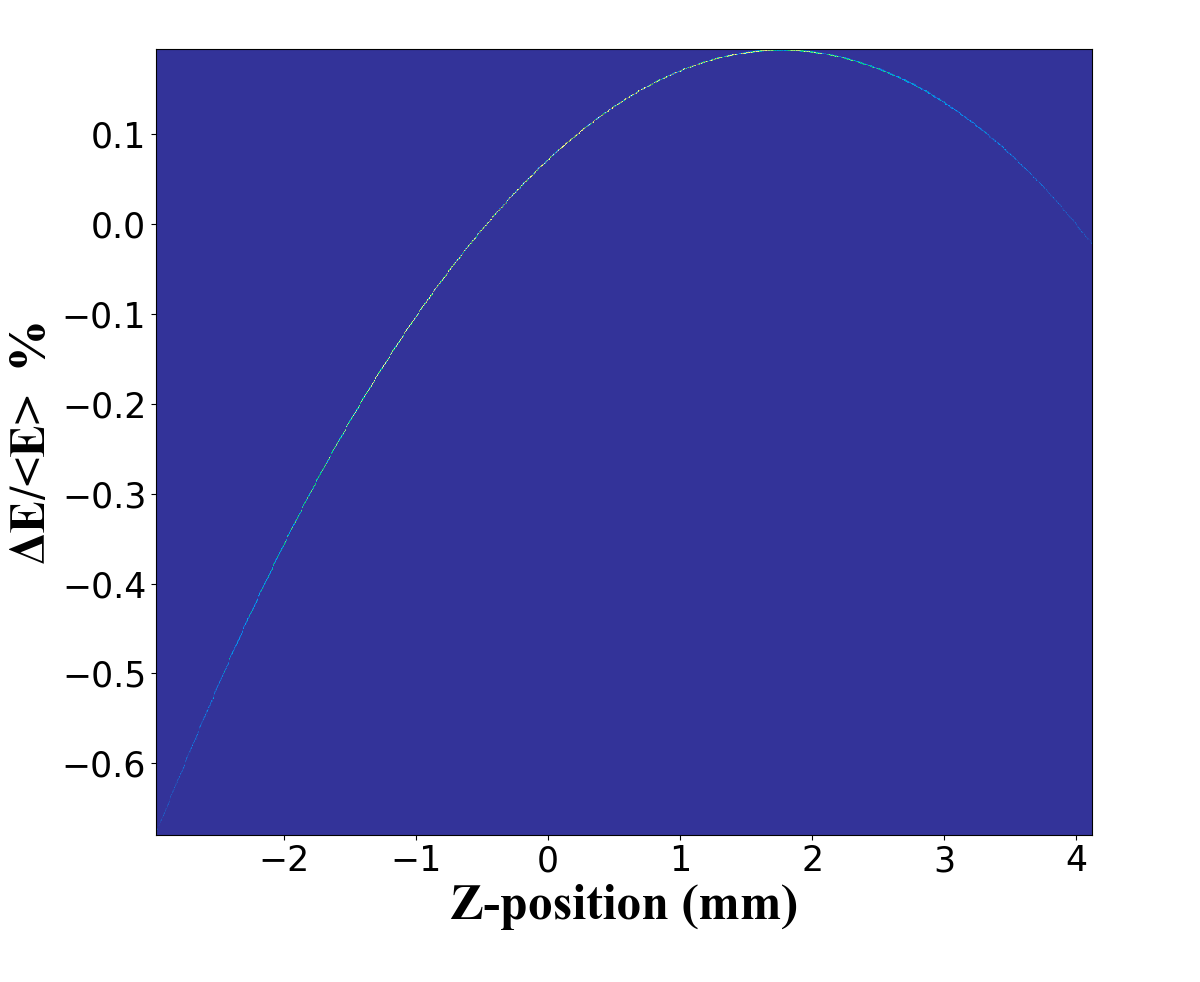}}
	\subfigure[]{
		\label{project4}	
		\includegraphics[width=0.48\linewidth]{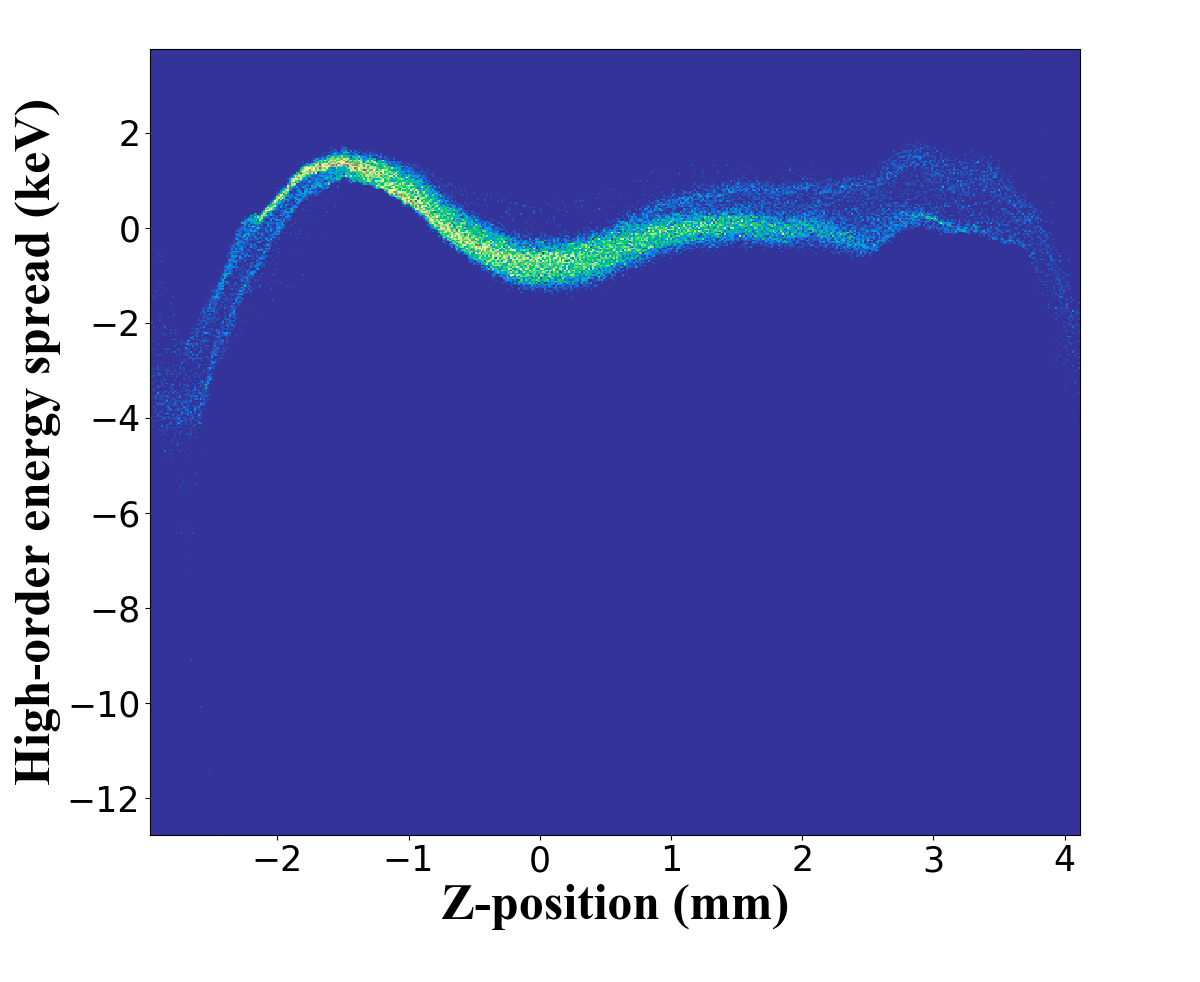}}
	\caption[width=1\textwidth]{Beam dynamics properties at the end of injector based on the selected solution from the last generation. (a) slice transverse normalized emittance; (b) current profile; (c) longitudinal phase space; (d)high order energy spread. The bunch head is to the left of the coordinate.}
	\label{property} 
\end{figure}

Eventually, the compromise is made and the beam properties are shown in the FIG.\ \ref{project} as the blue dots. The normalized transverse emittance is 0.223 mm·mrad, with bunch length of 1.49 mm. The bunch length is slightly longer compared with the previous design for mitigating the high-order terms value in the correlated energy spread to be smaller than 1 keV at the injector exit, which is much lower than previous design. Additionally, the slice transverse emittance, which is significant to FEL lasing performance, should be kept at an acceptable value. Detailed beam dynamics properties are presented in FIG.\ \ref{property}. 

\section{DISCUSSION AND CONCLUSION}

In this paper, an evolutionary many-objective optimization algorithm (NSGA-III) has been employed to optimize the physical design of a photoinjector section of a high-repetition-rate XFEL. In the optimization, the normalized emittance, bunch length, high-order correlated energy spread, and current profile skewness are treated as objectives. The NSGA-III not only optimizes the injector beam properties but also analyzes the relationship between different beam qualities based on the final Pareto front to select the optimal working point \cite{yan}. In the case of SHINE, the optimization results show that the smaller high order energy spread and more symmetric current profile can be achieved in the case of relatively waker velocity bunching which results in the slightly longer bunch length at the injector exit. The simulation results show that the property of high-order energy spread can be optimized to 0.87 keV, and the bunch length is 1.49 mm with normalized transverse emittance to be 0.223 mm·mrad. Additionally, the nonlinear density modulation in the buncher brings not only the nonlinear correlated energy distribution but also the asymmetric longitudinal distribution which generates the potential destructive implications for magnetic compression in the following beamline transport. Hence, the methods of longitudinal bunching compensation should be studied and conducted in the following work. 

\section{ACKNOWLEDGEMENTS}
This work was supported by the Youth Innovation Promotion Association CAS (2021282).

%
%
\ifboolexpr{bool{jacowbiblatex}}%
	{\printbibliography}%
	{%

} 
%
%

\end{document}